\documentclass[letterpaper, 10 pt, conference]{ieeeconf}

\IEEEoverridecommandlockouts
\usepackage{graphics} % for pdf, bitmapped graphics files
\usepackage{epsfig} % for postscript graphics files
\usepackage{mathptmx} % assumes new font selection scheme installed
\usepackage{times} % assumes new font selection scheme installed
\usepackage[fleqn]{amsmath} % assumes amsmath package installed
\usepackage{amssymb}  % assumes amsmath package installed

\usepackage{fancyhdr}
\usepackage{subfigure}
\usepackage{graphicx}
\usepackage{algorithmic}
\usepackage{algorithm}
\usepackage{mathrsfs}

\newtheorem{definition}{\hspace{1em}Definition}

\title{\LARGE \bf
Positioning Accuracy Improvement via Distributed Location Estimate in Cooperative Vehicular Networks}

\author{\authorblockN{Kai Liu,
Hock Beng Lim}
\authorblockA{Intelligent Systems Centre\\
Nanyang Technological University, Singapore
\\ Email: \{liukai, hblim\}@ntu.edu.sg}
}

\begin{document}

\maketitle
\thispagestyle{empty}
\pagestyle{empty}

\begin{abstract}
%\boldmath
The development of cooperative vehicle safety (CVS) applications, such as collision warnings, turning assistants, and speed advisories, etc., has received great attention in the past few years. Accurate vehicular localization is  essential to enable these applications. In this study, motivated by the proliferation of the Global Positioning System (GPS) devices, and the increasing sophistication of wireless communication technologies in vehicular networks, we propose a distributed location estimate algorithm to improve the positioning accuracy via cooperative inter-vehicle distance measurement. In particular, we compute the inter-vehicle distance based on raw GPS pseudorange measurements, instead of depending on traditional radio-based ranging techniques, which usually either suffer from  high hardware cost or have inadequate positioning accuracy. In addition, we improve the estimation of the vehicles' locations only based on the inaccurate GPS fixes, without using any anchors with known exact locations. The algorithm is decentralized, which enhances its practicability in highly dynamic vehicular networks. We have developed a simulation model to evaluate the performance of the proposed algorithm, and the results demonstrate that the algorithm can significantly improve the positioning accuracy.
\end{abstract}

\section{Introduction}

With recent advances in wireless communication technologies in vehicular networks, the cooperative vehicle safety (CVS) applications are expected to offer fundamental breakthroughs in enhancing the road safety. Accurate vehicle positioning is vital to enabling  many safety-critical applications, such as cooperative collision warnings, turning assistants, and speed advisories, to name but a few. In these applications, the positioning has to be accurate enough such that the vehicles could be differentiated at the lane level. However, current commercially available GPS devices, which are widely used in ground vehicles for positioning and navigation, typically report errors on the order of 10 meters (\cite{boukerche2008vehicular, schubert2007accurate}). Thus, it is difficult to recover the lane level relationship among vehicles within a vicinity region.

Much effort has been devoted to increasing the positioning accuracy in vehicular networks. Leveraging reference points (i.e. anchors or base stations) with known exact positions is one of the most prevalent solutions to correct the positioning error. Differential GPS (DGPS)  and assisted GPS (A-GPS)  are two of the well-known techniques relying on ground-based reference points to improve the GPS accuracy. Many localization systems in vehicular networks have employed similar concepts \cite{du2008next}. R. W. Ouyang et al. \cite{ouyang2010gps} utilized the time of arrival (TOA) measurement taken from terrestrial reference stations to improve the initial GPS fix accuracy. Three algorithms, including a geometric approach, a weighted least squares based approach and a closed-form approach, are proposed to enhance the positioning performance under different driving conditions. Inspired by the concept of DGPS, E.K Lee et al. \cite{lee2009rf} proposed a mobile version of DGPS system, which exploits the radio-frequency identification (RFID) technique in vehicle positioning. In the proposed system, when a vehicle obtains a known precise position from an RFID tag installed along the road via its RFID reader, this vehicle would act as a moving reference point temporarily by computing the GPS errors, and then sharing the errors with its neighbors to help them correct their GPS coordinates. Although most of the localization systems based on reference points have been demonstrated effective, they require the support of particular hardwares or large-scale infrastructures, while some of them are not yet globally available.

Currently, GPS becomes a mature and widely applied localization system. The proliferation of GPS makes it the de facto solution for vehicle navigation and localization, though it still suffers from some issues such as accuracy, reliability, and robustness to varying degrees. Meanwhile, the increasing sophistication of wireless communication technologies in vehicular networks, such as the dedicated short range communications (DSRC), encourages the development of cooperative applications \cite{liu2012adaptive}. Motivated by the above observations, in this work, we are dedicated to improving the positioning accuracy based on GPS pseudorange measurements in a cooperative vehicular network.

The main contributions of this work are summarized as follows. First, existing solutions which utilize inter-vehicle distances in improving the positioning accuracy (\cite{parker2006vehicle, ?apkun2002gps, drawil2008vehicular, parker2007cooperative}), in one way or another, rely on the radio-based ranging technologies, such as Time of Arrival (ToA) \cite{ouyang2010gps}, Time Difference of Arrival (TDoA) \cite{bard1999time}, Angle-of-arrival (AOA) \cite{rong2006angle}, and  Received Signal Strength (RSS) \cite{parker2007vehicular}. Typically, these technologies either require particular ranging sensors causing a high hardware cost, or suffer from inadequate ranging accuracy  \cite{alam2010positioning}. In contrast, the proposed algorithm is designed to compute  inter-vehicle distances with high accuracy only by the acquisition of GPS information without using ranging sensors. Second, the proposed algorithm estimates the vehicle location solely based on inaccurate GPS measurements without the assumption of any reference points. Last, in view of the highly dynamic nature in vehicular networks, the proposed algorithm is operated in a distributed fashion to make it a potentially practical solution. Note that this study focuses on improving the positioning accuracy in open areas with reasonably good GPS signal (i.e. at least four satellites available for localization), such as highways, suburban areas, and rural areas, etc. Nevertheless, in other environments,  solutions for compensating GPS outages and overcoming GPS limitations can be referred to (\cite{parker2006vehicle, ?apkun2002gps, drawil2008vehicular, drawil2009toward}).

The rest of this paper is organized as follows. Section \ref{pre} reviews the inter-vehicle distance detection technique and formulates the research problem. Section \ref{alg}  proposes a distributed location estimate algorithm. Section \ref{pe} develops the simulation model and gives the performance evaluation. Last, we conclude this study and outline the further work in Section \ref{con}.

\section{Preliminary}\label{pre}
\subsection{GPS pseudorange based distance measurement}
Many studies on cooperative vehicular localization assume the availability of the distance between two neighboring vehicles to improve the positioning accuracy. Typical techniques assumed to obtain the inter-vehicle distance include ToA, TDoA, AOA, and RSS, which are the radio-based ranging techniques proposed in cellular networks. In vehicular networks, however, these techniques can hardly achieve the ranging accuracy required by  safety-critical applications \cite{alam2010positioning}. In order to improve the accuracy of inter-vehicle distance measurement, we have proposed an algorithm based on the sharing of GPS pseudorange information \cite{yang2012vehicle}. In this section, we recapitulate the key ideas of the algorithm, which forms the basis of this study.

In the GPS system, the distance from a vehicle to each satellite is derived from an estimated time of transmission, and this distance is called the \emph{pseudorange}. Several sources contribute to the errors in the pseudorange measurement. The pseudorange ($PR_a^i$) between GPS receiver $a$  to satellite $i$  can be decomposed into \cite{kaplan2006understanding}:
\begin{eqnarray}
PR_a^i  = R_a^i  + t_a  + x^i  + \varepsilon _a^i
\end{eqnarray}
where $R_a^i$ is the true distance between satellite $i$  and receiver $a$;  $t_a$ is the error caused by receiver  $a$'s clock bias;  $x^i$ is the common noise related to satellite $i$  that are shared by each GPS receiver within a vicinity region, including satellite clock bias,  atmospheric delay, and errors in the broadcasted ephemeris;  $\varepsilon_a^i$ is the non-common noise specific to receiver $a$  and satellite $i$ , including the multipath and code acquisition noises. By taking the difference between the pseudoranges of the two receivers $a$ and $b$  to the same satellite $i$ , the common noise due to satellite $i$ can be effectively removed:
\begin{eqnarray}
S^i_{ab} &=& PR^i_a - PR^i_b  \nonumber \\
        &=& (R^i_a - R^i_b) + (t_a - t_b)+ (\varepsilon^i_a - \varepsilon^i_b)   \nonumber \\
        &=& \Delta R^i_{ab} + (t_a - t_b)+ (\varepsilon^i_a - \varepsilon^i_b)
\end{eqnarray}

\begin{figure}
\centering
\includegraphics[width=2.5in]{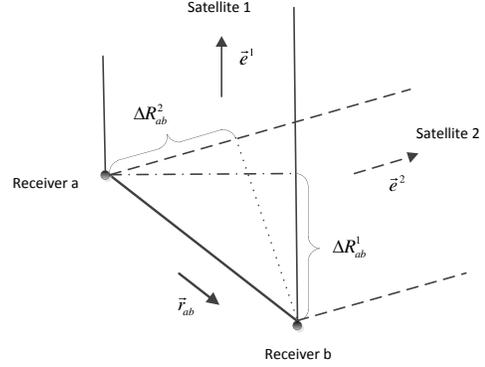}
\caption{Pseudorange double difference}
\label{fig:1}
\end{figure}
\noindent where $S^i_{ab}$ is the single difference of pseudorange measurements, and $\Delta R^i_{ab}$ is the difference between the true ranges from receiver $a$ and $b$   to satellite $i$ . As the true ranges from  satellite $i$ to receivers $a$  and $b$   are much larger than the distance between  $a$ and $b$ , the two vectors pointing from $i$  to $a$ and $b$  are nearly parallel to each other. As illustrated in Figure \ref{fig:1}, $\Delta R^i_{ab}$  can thus be approximated by:
\begin{eqnarray}
\Delta R_{ab}^i  = \vec e^i  \cdot \vec r_{ab}
\end{eqnarray}
where $\vec e^i$ is the unit vector pointing from receiver $a$ (or $b$) to satellite $i$, and $\vec r_{ab}$ is the distance vector between the receivers $a$ and $b$. When double difference is used, the clock bias of the receivers  $a$ and $b$  can be further removed.
\begin{eqnarray}
D^{ij}_{ab}    &=& S^i_{ab}-S^j_{ab}   \nonumber \\
                &=& [\Delta R^i_{ab} - \Delta R^j_{ab}]+ [(\varepsilon^i_a - \varepsilon^i_b)- (\varepsilon^j_a - \varepsilon^j_b)]   \nonumber \\
                &=& [\vec{e}^i - \vec{e}^j]\cdot \vec{r}_{ab} + [(\varepsilon^i_a - \varepsilon^i_b)- (\varepsilon^j_a - \varepsilon^j_b)]
\label{eq:dd}
\end{eqnarray}

Given a set of satellites $\{ 0,1,...,n\}$ shared by the receivers $a$ and $b$, Eq.\ref{eq:dd}  can be reorganized into:
\begin{eqnarray}
\mathbf{D}_{ab}= \mathbf{H} \vec{r}_{ab} + \boldsymbol{\epsilon}
\label{eq:dd_vec}
\end{eqnarray}
where $\mathbf{D}_{ab}$ is the column vector of pseudorange double differences with respect to the satellites $0$ and $i$ ($1 \le i \le n$), and $\mathbf{D}_{ab}=[D^{10}_{ab} \; D^{20}_{ab} \; \cdots \; D^{n0}_{ab}]^T$. Accordingly, $\mathbf{H}$ is the column vector of the difference between two unit vectors, and $\mathbf{H} = [(\vec{e}^1 - \vec{e}^0) \; (\vec{e}^2 - \vec{e}^0) \; \cdots \; (\vec{e}^n - \vec{e}^0)]^T$.  $\boldsymbol{\epsilon}$ is the column vector of aggregated non-common noises, and  $\boldsymbol{\epsilon} = [\left( {(\varepsilon _a^1  - \varepsilon _b^1 ) - (\varepsilon _a^0  - \varepsilon _b^0 )} \right)...\left( {(\varepsilon _a^n  - \varepsilon _b^n ) - (\varepsilon _a^0  - \varepsilon _b^0 )} \right)]^T$. Assume $\boldsymbol{\epsilon}$ is zero mean and equal variance, $\vec{r}_{ab}$ can be approximately solved by the linear least squares estimator:
\begin{eqnarray}
\vec{r}_{ab}= (\mathbf{H}^T \mathbf{H})^{-1}  \mathbf{H}^T  \mathbf{D}_{ab}
\end{eqnarray}

In order to further improve the accuracy of the computed distance $\vec{r}_{ab}$, we have taken the Carrier to Noise Ratio (CNR) of raw pseudorange measurements into account for noise mitigation, and proposed an algorithm called  \emph{weighted least squares double difference }(WLS-DD). The final form of the weight matrix $\mathbf{W}$ is represented by:
\begin{eqnarray}
\mathbf{W} = diag\left(\frac{(\phi^1_a)^2 \cdot (\phi^1_b)^2}{(\phi^1_a)^2 + (\phi^1_b)^2}, \cdots, \frac{(\phi^n_a)^2 \cdot (\phi^n_b)^2}{(\phi^n_a)^2 + (\phi^n_b)^2}\right)
\end{eqnarray}
where $diag(\cdot)$  denotes a diagonal matrix, and $\phi^i_a$ is the CNR value of the received signal from  satellite $i$ to receiver $a$. The detailed designing rationale of the weight matrix $\mathbf{W}$ can be referred to \cite{yang2012vehicle}. Finally, the distance between $a$  and $b$ is computed by:
\begin{eqnarray}
\vec{r}_{ab}= (\mathbf{H}^T \mathbf{W} \mathbf{H})^{-1}  \mathbf{H}^T \mathbf{W} \mathbf{D}_{ab}
\label{eq:wls}
\end{eqnarray}

We have demonstrated that the WLS-DD significantly increases the accuracy of the distance measurement via an extensive field testing. On the basis of the accurate distance measurement, we formulate the research problem of this study as follows.

\subsection{Problem statement}
The set of vehicles is denoted by $V = \{ v_1 ,v_2 ,...,v_{\left| V \right|} \}$, where $|V|$ is the total number of vehicles. The true location of vehicle $v_i$ ($1 \leq i\leq|V|$) is represented by the vector $\mathbf{v}_i$. $v_j$ is considered as the \emph{neighbor} of $v_i$ if the two vehicles are within their communication range. The set of $v_i$'s neighbors is denoted by $N_i$ ($N_i  \subset V$). For $\forall v_j  \in N_i $, the true distance between $v_i$ and $v_j$  is denoted by $d_{ij}$ ($d_{ij}  = \left\| {{\mathbf{v}_i}  - {\mathbf{v}}_j } \right\|$), while the computed distance (by WLS-DD) between $v_i$ and $v_j$  is denoted by $\tilde d_{ij} $. Denote the original GPS fix of $v_i$  as ${\mathbf{\tilde v}}_i$, which is typically  subject to the positioning error on the order of 10 meters (\cite{boukerche2008vehicular, schubert2007accurate}). Note that the accuracy of the  inter-vehicle distance computed by WLS-DD is much higher than that directly obtained from the original GPS fixes of the two neighboring vehicles, which is $\left\| {{\mathbf{\tilde v}}_i  - {\mathbf{\tilde v}}_j } \right\|$.

The overall objective of this work is to improve the accuracy of vehicular localization by leveraging the original GPS fixes and the computed inter-vehicle distances. Specifically, given a set of vehicles $V$ with their original GPS fixes ${\mathbf{\tilde v}} = \{ {\mathbf{\tilde v}}_1 ,{\mathbf{\tilde v}}_2 ,...,{\mathbf{\tilde v}}_{|V|} \} $, and the computed distance $\tilde d_{ij} $  between any two neighboring vehicles $v_i$ and $v_j$ ($v_i \in V$ and $v_j \in N_i$), the algorithm is committed to computing a set of estimated locations ${\mathbf{\hat v}} = \{ {\mathbf{\hat v}}_1 ,{\mathbf{\hat v}}_2 ,...,{\mathbf{\hat v}}_{|V|} \} $, so that the estimated location ${\mathbf{\hat v}}_i $ can approximate to the true location  ${\mathbf{v}}_i $. In brief, the problem is to minimize the value of $\sum\limits_{i = 1}^{|V|} {\left\| {{\mathbf{\hat v}}_i  - {\mathbf{v}}_i } \right\|} $, which is the overall error of the estimated locations. It is worth noting that, we aim to improve the positioning accuracy solely based on the inaccurate raw GPS measurements without the assumption of any particular equipments or infrastructures (i.e. ranging sensors for distance measurement, or anchors for error correction). The primary notations used in this work are summarized in Table \ref{table:notations}.

\begin {table}\footnotesize
\renewcommand{\arraystretch}{1.4}
\centering \caption{Summary of notations} \label{table:notations}
\begin{tabular}{|l|l|l|l}
\cline{1-3}
\multicolumn{1}{|c|}{\textbf{Notations}} & \multicolumn{1}{c|}{\textbf{Descriptions}} & \multicolumn{1}{c|}{\textbf{Notes}} &  \\
\cline{1-3}
\multicolumn{1}{|c|}{$V$} & \multicolumn{1}{c|}{total set of vehicles} & \multicolumn{1}{c|}{$V = \{ v_1 ,v_2 ,...,v_{\left| V \right|} \}$} &  \\
\cline{1-3}
\multicolumn{1}{|c|}{$N_i$} & \multicolumn{1}{c|}{neighbor set of $v_i$} & \multicolumn{1}{c|}{$N_i \subset V$} &  \\
\cline{1-3}
\multicolumn{1}{|c|}{$V^k$} & \multicolumn{1}{c|}{subset of vehicles with } & \multicolumn{1}{c|}{$V^k=N_k+\{v_k\}$} &  \\
\multicolumn{1}{|c|}{} & \multicolumn{1}{c|}{the pivot $v_k$} & \multicolumn{1}{c|}{and $V^k \subseteq V$} &  \\
\cline{1-3}
\multicolumn{1}{|c|}{$\mathbf{v}_i$} & \multicolumn{1}{c|}{true location of $v_i$} & \multicolumn{1}{c|}{} &  \\
\cline{1-3}
\multicolumn{1}{|c|}{$d_{ij}$} & \multicolumn{1}{c|}{true distance between } & \multicolumn{1}{c|}{$d_{ij}  = \left\| {{\mathbf{v}}_i  - {\mathbf{v}}_j } \right\|$
} &  \\
\multicolumn{1}{|c|}{} & \multicolumn{1}{c|}{$v_i$ and $v_j$} & \multicolumn{1}{c|}{} &  \\
\cline{1-3}
\multicolumn{1}{|c|}{${\mathbf{\tilde v}}_i $} & \multicolumn{1}{c|}{original GPS fix of $v_i$} & \multicolumn{1}{c|}{} &  \\
\cline{1-3}
\multicolumn{1}{|c|}{$\tilde d_{ij} $} & \multicolumn{1}{c|}{computed distance between } & \multicolumn{1}{c|}{computed by WLS-DD} &  \\
\multicolumn{1}{|c|}{} & \multicolumn{1}{c|}{ $v_i$ and $v_j$} & \multicolumn{1}{c|}{} &  \\
\cline{1-3}
\multicolumn{1}{|c|}{${\mathbf{\delta }}_i $} & \multicolumn{1}{c|}{GPS error of $v_i$} & \multicolumn{1}{c|}{${\mathbf{\delta }}_i  = {\mathbf{v}}_i  - {\mathbf{\tilde v}}_i $} &  \\
\cline{1-3}
\multicolumn{1}{|c|}{${\mathbf{\hat v}}_i^k $} & \multicolumn{1}{c|}{tentative estimated location } & \multicolumn{1}{c|}{} &  \\
\multicolumn{1}{|c|}{} & \multicolumn{1}{c|}{of $v_i$ in subset $V^k$} & \multicolumn{1}{c|}{} &  \\
\cline{1-3}
\multicolumn{1}{|c|}{${\mathbf{\hat v}}_i $} & \multicolumn{1}{c|}{final estimated location of $v_i$} & \multicolumn{1}{c|}{} &  \\
\cline{1-3}
\multicolumn{1}{|c|}{${\mathbf{\hat V}}^k $} & \multicolumn{1}{c|}{set of tentative estimated } & \multicolumn{1}{c|}{} &  \\
\multicolumn{1}{|c|}{} & \multicolumn{1}{c|}{locations computed by $v_k$} & \multicolumn{1}{c|}{} &  \\
\cline{1-3}
\multicolumn{1}{|c|}{$w_i$} & \multicolumn{1}{c|}{weight of the pivot $v_i$} & \multicolumn{1}{c|}{} &  \\
\cline{1-3}
\end{tabular}
\end{table}
\normalsize

\section{A Distributed Location Estimate Algorithm} \label{alg}
In this section, we propose a distributed location estimate algorithm (DLEA) to improve the localization accuracy in a cooperative vehicular network. The detailed procedures of  DLEA along with its designing rationale are presented below.

\subsection{Tentative location estimate}
A prerequisite to enable the cooperative localization is that, for any vehicle $v_i$ ($v_i \in V$), there is at least one neighboring vehicle $v_j$ ($v_j \in V$),  namely, $N_i  \ne \emptyset $ for $\forall v_i  \in V$. With this prerequisite, each vehicle $v_i$ is able to estimate the location based on the information (i.e., computed distances, and original GPS fixes) received from its neighbors. To facilitate the implementation of the distributed location estimate, we define the \emph{subset of vehicles} as follows.

\begin{definition}\label{def1}
\textbf{subset of vehicles} Given a vehicle $v_k$ ($v_k \in V$), it is considered as a pivot of a subset  $V^k$, when $V^k$ is comprised of $v_k$ and all of its neighbors $\{ v_j |v_j  \in N_k \}$, namely,  $V^k  = N_k  + \{ v_k \}$.
\end{definition}

Note that each vehicle $v_k$ will act as a pivot and construct a corresponding subset $V^k$. In the following, we transform the tentative location estimate into a constrained non-linear optimization problem. In this way, each pivot vehicle $v_k$ will compute a set of \emph{tentative estimated locations}, which is represented by ${\mathbf{\hat V}}^k  = \{ {\mathbf{\hat v}}_i^k |v_i  \in V^k \} $, where ${\mathbf{\hat v}}_i^k $ is the tentative estimated location for $v_i$ ($v_i \in V^k$).

Given a vehicle $v_i$ ($v_i \in V^k$) with its original GPS fix ${\mathbf{\tilde v}}_i $, the GPS error ${\mathbf{\delta }}_i $ can be represented by ${\mathbf{\delta }}_i  = {\mathbf{v}}_i  - {\mathbf{\tilde v}}_i $, where $\mathbf{v}_i$ is the true location of $v_i$. Since the GPS error is caused by many independent sources, such as satellite clock bias, atmospheric delay, acquisition noises , and multipath, etc., it is commonly to assume that the error $\mathbf{\delta}$ follows the Gaussian distribution (\cite{parker2006vehicle,drawil2009toward}), namely,  ${\mathbf{\delta }} \sim N({\mathbf{\mu }},{\mathbf{\sigma }}^2 )$, where ${\mathbf{\mu }}$ is the mean, and ${\bf{\sigma }}^2 $ is the variance. Denote the probability density function as $\varphi ({\mathbf{\delta }})$, then $\varphi ({\bf{\delta }}) = \frac{1}{{\sqrt {2\pi } {\bf{\sigma }}}} \cdot e^{ - \frac{{({\bf{\delta }} - {\bf{\mu }})^2 }}{{2{\bf{\sigma }}^2 }}}$. Given the  $\varphi ({\mathbf{\delta }})$,  the value of the error  (${\mathbf{v}}_i  - {\mathbf{\tilde v}}_i $) is most likely to distribute around the point where the highest probability density is achieved. Therefore, in order to maximize the possibility that the tentative estimated location (${\mathbf{\hat v}}_i^k $) would approximate to the true location (${\mathbf{v}}_i $), it is expected to maximize the function of $\varphi ({\mathbf{\hat v}}_i^k  - {\mathbf{\tilde v}}_i)$. Besides, the tentative estimated locations should also satisfy the distance constraint between any two neighboring vehicles, which is represented by: $\left\| {{\mathbf{\hat v}}_j^k  - {\mathbf{\hat v}}_k^k } \right\| = \tilde d_{jk} $, for  $\forall v_j \in N_k$, where $\tilde d_{jk}$ is the computed distance between the pivot $v_k$ and its neighbor $v_j$. In practice, the road space constraint (e.g.  boundary coordinates), if applicable, can be also included to enhance the positioning accuracy. To sum up, the set of tentative estimated locations ${\bf{\hat V}}^k $ computed by the pivot  $v_k$ can be derived from:
\begin{eqnarray}
{\mathbf{\hat V}}^k  = arg{\rm{ }}\mathop {\max }\limits_{{\mathbf{\hat V}}^k } \sum\limits_{v_i  \in V^k } {\varphi ({\mathbf{\hat v}}_i^k  - {\mathbf{\tilde v}}_i )}
\end{eqnarray}
subject to that
\begin{align*}
&\left\| {{\mathbf{\hat v}}_j^k  - {\mathbf{\hat v}}_k^k } \right\| = \tilde d_{jk}   \quad\mbox{for}\quad \forall v_j  \in N_k,~and\\&  \,\,\,\,
{\mathbf{\hat v}}_i^k  \in \mathscr{S}  \qquad\qquad\mbox{for}\quad \forall v_i  \in V^k.
\end{align*}
where\\
$V^k$ is the subset with the pivot vehicle $v_k$;\\
$\mathbf{\hat v}_i^k$ is the tentative estimated location for $v_i$ computed by $v_k$;\\
$\mathbf{\tilde v}_i$ is the original GPS fix of $v_i$;\\
$\tilde d_{jk}$ is the computed distance between $v_j$ and $v_k$;\\
$N_k$ is the neighbor set of $v_k$;\\
$\mathscr{S}$ represents the road space constraint.

\subsection{An example}
\begin{figure}
\centering
\includegraphics[width=2.6in]{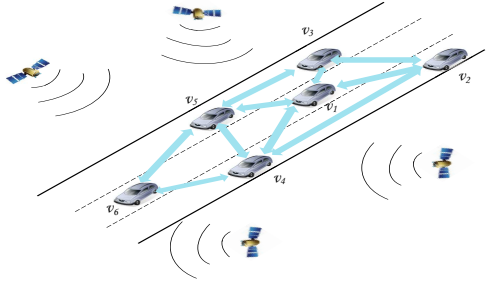}
\caption{An example}
\label{fig:2}
\end{figure}

We illustrate the computing details of for the tentative estimated locations  by the following example. As shown in Figure \ref{fig:2}, given the road space $\mathscr{S}$, the total set of vehicles is $V = \{ v_1 ,v_2 ,v_3 ,v_4 ,v_5 ,v_6 \}$. The edge with double arrows  represents the two vehicles are within their communication range. In other words, they can share the GPS pseudorange measurement with each other and compute the inter-vehicle distance. Taken   $v_1$ as an example, the neighbor set of $v_1$ is represented by $N_1  = \{ v_2 ,v_3 ,v_4 ,v_5 \}$, and the subset of vehicles with pivot $v_1$  is  $V^1  = \{ v_1 ,v_2 ,v_3 ,v_4 ,v_5 \}$. The set of tentative estimated locations ${\mathbf{\hat V}}^1  = \{ {\mathbf{\hat v}}_1^1 ,{\mathbf{\hat v}}_2^1 ,{\mathbf{\hat v}}_3^1 ,{\mathbf{\hat v}}_4^1 ,{\mathbf{\hat v}}_5^1 \} $ is computed by:
\setlength{\abovedisplayskip}{2mm}
\setlength{\belowdisplayskip}{2mm}
\begin{eqnarray}
{\mathbf{\hat V}}^1  = arg{\rm{ }}\mathop {\max }\limits_{{\mathbf{\hat V}}^1 } \left( {\varphi ({\mathbf{\hat v}}_1^1  - {\mathbf{\tilde v}}_1 ) + \varphi ({\mathbf{\hat v}}_2^1  - {\mathbf{\tilde v}}_2 )... + \varphi ({\mathbf{\hat v}}_5^1  - {\mathbf{\tilde v}}_5 )} \right)
\end{eqnarray}
Subject to that
\begin{align*}
&\left\| {{\mathbf{\hat v}}_2^1  - {\mathbf{\hat v}}_1^1 } \right\| = \tilde d_{21},
\left\| {{\mathbf{\hat v}}_3^1  - {\mathbf{\hat v}}_1^1 } \right\| = \tilde d_{31},
\left\| {{\mathbf{\hat v}}_4^1  - {\mathbf{\hat v}}_1^1 } \right\| = \tilde d_{41}, \\&
\left\| {{\mathbf{\hat v}}_5^1  - {\mathbf{\hat v}}_1^1 }\right\| = \tilde d_{51}, \quad and \quad {\mathbf{\hat v}}_i^1  \in \mathscr{S}  \quad \mbox{for}\quad \forall v_i  \in V^1.
\end{align*}

To solve the above constrained non-linear optimization problem, for clear exposition, we consider in a two-dimension space. Nevertheless, note that it is straightforward to extend the solution into a three-dimension space. In an $x$-$y$ coordinate, the tentative estimated location ${\mathbf{\hat v}}_i^k $, the GPS fix ${\mathbf{\tilde v}}_i $, and the error ${\mathbf{\delta }}_i$ can be expanded to: ${\mathbf{\hat v}}_i^k (\hat v_{i_x }^k ,\hat v_{i_y }^k )$, ${\mathbf{\tilde v}}_i (\tilde v_{i_x } ,\tilde v_{i_y } )$, and ${\bf{\delta }}_i (\delta _{i_x } ,\delta _{i_y } )$, respectively. Accordingly, the objective function is decomposed into $\varphi (\hat v_{i_x }^k  - \tilde v_{i_x } ) \cdot \varphi (\hat v_{i_y }^k  - \tilde v_{i_y } )$. Given the mean value ${\mathbf{\mu }}(\mu _x ,\mu _y )$, and the standard deviation ${\bf{\sigma }}(\sigma _x ,\sigma _y )$, we have:
\begin{eqnarray}
\varphi (\hat v_{i_x }^k  - \tilde v_{i_x } ) &=& \frac{1}{{\sqrt {2\pi } \sigma _x }} \cdot e^{ - \frac{{(\hat v_{i_x }^k  - \tilde v_{i_x }  - \mu _x )^2 }}{{2\sigma _x ^2 }}}\nonumber ~~and\\
\varphi (\hat v_{i_y }^k  - \tilde v_{i_y } ) &=& \frac{1}{{\sqrt {2\pi } \sigma _y }} \cdot e^{ - \frac{{(\hat v_{i_y }^k  - \tilde v_{i_y }  - \mu _y )^2 }}{{2\sigma _y ^2 }}}
\label{eq:pdf}
\end{eqnarray}

\noindent The set of tentative estimated locations is computed by:
\begin{align}
{\mathbf{\hat V}}^k  = arg{\rm{ }}\mathop {\max }\limits_{{\mathbf{\hat V}}^k } \sum\limits_{v_i  \in V^k } {\varphi (\hat v_{i_x }^k  - \tilde v_{i_x } ) \cdot \varphi (\hat v_{i_y }^k  - \tilde v_{i_y } )}
\label{eq:example}
\end{align}
According to Eq.\ref{eq:pdf}, let
\begin{eqnarray*}
 M &=& \varphi (\hat v_{i_x }^k  - \tilde v_{i_x } ) \cdot \varphi (\hat v_{i_y }^k  - \tilde v_{i_y } ) \\
 {\rm{    }} &=& \frac{1}{{\sqrt {2\pi } \sigma _x }} \cdot e^{ - \frac{{(\hat v_{i_x }^k  - \tilde v_{i_x }  - \mu _x )^2 }}{{2\sigma _x ^2 }}}  \cdot \frac{1}{{\sqrt {2\pi } \sigma _y }} \cdot e^{ - \frac{{(\hat v_{i_y }^k  - \tilde v_{i_y }  - \mu _y )^2 }}{{2\sigma _y ^2 }}}
\end{eqnarray*}
then
\begin{align}
\log M =  -A+B \nonumber
\end{align}
\noindent where
\begin{align}
A =   {\frac{{(\hat v_{i_x }^k  - \tilde v_{i_x }  - \mu _x )^2 }}{{2\sigma _x ^2 }} + \frac{{(\hat v_{i_y }^k  - \tilde v_{i_y }  - \mu _y )^2 }}{{2\sigma _y ^2 }}} \nonumber
\end{align}
\noindent and
\begin{align}
B =  \log \frac{1}{{\sqrt {2\pi } \sigma _x }} + \log \frac{1}{{\sqrt {2\pi } \sigma _y }}\nonumber
\end{align}
\noindent Since $B$ is a constant, the objective of maximizing the value of $M$ is equivalent to the objective of minimizing the value of $A$. Accordingly, Eq.\ref{eq:example} can be transformed to:
\begin{eqnarray}
{\mathbf{\hat V}}^k  = arg{\rm{ }}\mathop {\min }\limits_{{\bf{\hat V}}^k } \sum\limits_{v_i  \in V^k } {\frac{{(\hat v_{i_x }^k  - \tilde v_{i_x }  - \mu _x )^2 }}{{2\sigma _x ^2 }} + \frac{{(\hat v_{i_y }^k  - \tilde v_{i_y }  - \mu _y )^2 }}{{2\sigma _y ^2 }}}
\end{eqnarray}
subject to that
\begin{align*}
&\sqrt {(\hat v_{j_x }^k  - \hat v_{k_x }^k )^2  + (\hat v_{j_y }^k  - \hat v_{k_y }^k )^2 }  = \tilde d_{jk} \quad\mbox{for}\quad
\forall v_j  \in N_k, ~and \\&
S_{lb}^x  \le \hat v_{i_x }^k  \le S_{ub}^x, \quad S_{lb}^y  \le \hat v_{i_y }^k  \le S_{ub}^y \,\,\,\,\quad\mbox{for}\quad
\forall v_i  \in V^k.
\end{align*}
where
$S_{lb}^x ,S_{ub}^x ,S_{lb}^y$, and $S_{ub}^y $ represent the lower bound and the upper bound of the $x$ and $y$ coordinates confined by the road space, respectively.

\subsection{Final location estimate}

With the above method, each vehicle $v_k$ ($v_k \in V$), when acting as a pivot, will compute in total of $|V^k|$ tentative estimated locations, including ${\mathbf{\hat v}}_k^k $ (for itself) and a set $\{ {\mathbf{\hat v}}_i^k |v_i  \in N_k \} $ for each of its neighbors ($|V^k | = |N_k | + 1$).  This implies that, each vehicle $v_k$ will also obtain in total of  $|V^k|$   tentative estimated locations, including ${\mathbf{\hat v}}_k^k $  computed by itself, and a set $\{ {\mathbf{\hat v}}_k^i |v_i  \in N_k \} $ computed by each of its neighbor $v_i$  (when $v_i$ acting as a pivot). Intuitively, given a pivot $v_k$, the more vehicles are within the subset $V^k$, the higher possibility that the computed ${\mathbf{\hat v}}_i^k $
 is closer to the true location of $v_i$, as there are more constraints (i.e., distances between two neighboring vehicles) are applied in position estimate. In view of this, we define the \emph{weight of a pivot vehicle} as follows.

\vspace{0.5em}
\begin{definition}\label{def2}
\textbf{weight of a pivot vehicle} Given a pivot $v_k$  with its corresponding subset $V^k$, the weight of $v_k$, denoted by $w_k$, is defined as the number of vehicles in $V^k$, namely, $w_k  = \left| {V^k } \right|$.
\end{definition}

The final estimated location ${\mathbf{\hat v}}_k $ is a weighted average of all the tentative estimated locations for $v_k$ ($\{ {\mathbf{\hat v}}_k^k \}  + \{ {\mathbf{\hat v}}_k^i |v_i  \in N_k \} $), and the set is equivalent to  $\{ {\mathbf{\hat v}}_k^s |v_s  \in V^k \} $, where ${\mathbf{\hat v}}_k^s $ is the tentative estimated location for $v_k$ computed by the pivot $v_s$. The weight of ${\bf{\hat v}}_k^s $ is defined as follows.

\vspace{0.5em}
\begin{definition}\label{def3}
\textbf{weight of a tentative estimated location} Given a tentative estimated location for $v_k$ computed by the pivot $v_s$ (${\mathbf{\hat v}}_k^s $), the weight of ${\mathbf{\hat v}}_k^s $  is defined as the weight of $v_s$ ($w_s$) over the sum of the weight of each $v_i$ in the subset $V^k$, which is calculated by $\frac{{w_s }}{{\sum\limits_{v_i  \in V^k } {w_i } }}$.
\end{definition}

With the above definitions, the final estimated location ${\bf{\hat v}}_k $ is a linear combination of all the tentative estimated locations ${\bf{\hat v}}_k^s $, which is computed by:
\begin{eqnarray}
{\mathbf{\hat v}}_k  = \sum\limits_{v_s  \in V^k } {\frac{{w_s }}{{\sum\limits_{v_i  \in V^k } {w_i } }} \cdot {\mathbf{\hat v}}_k^s }
\end{eqnarray}

Taken  $v_6$ shown in Figure \ref{fig:2} as an example, the subset $V^6  = \{ v_4 ,v_5 ,v_6 \}$. Therefore, $v_6$ will get three tentative estimated locations, ${\mathbf{\hat v}}_6^4 $,  ${\mathbf{\hat v}}_6^5$, and ${\mathbf{\hat v}}_6^6 $, which are computed by $v_4$, $v_5$, and $v_6$, respectively. Besides, the weight of each pivot vehicle is $w_4=4$, $w_5=4$, and $w_6=2$. So, the final estimated location is computed by:
\begin{align*}
{\mathbf{\hat v}}_6  = \frac{1}{{w_4  + w_5  + w_6 }} \cdot (w_4 {\mathbf{\hat v}}_6^4  + w_5 {\mathbf{\hat v}}_6^5  + w_6 {\mathbf{\hat v}}_6^6 )
\end{align*}

The main procedures of  DLEA  is summarized as follows.

\textbf{Step 1:} To compute the distance between any two neighboring vehicles based on WLS-DD.

\vspace{0.3em}

\textbf{Step 2:} Each vehicle $v_i$ shares its original GPS fix ${\mathbf{\tilde v}}_i $ with its neighbors.

\vspace{0.3em}

\textbf{Step 3:} Each vehicle $v_i$ acts as a pivot and computes a set of tentative estimated locations ${\mathbf{\hat V}}^i $.

\vspace{0.3em}

\textbf{Step 4:} Each vehicle  $v_i$ shares the computed ${\mathbf{\hat V}}^i $ with its neighbors.

\vspace{0.3em}

\textbf{Step 5:} Each vehicle $v_i$  computes its final estimated location ${\mathbf{\hat v}}_i $ based on all of its tentative estimated locations.

\section{Performance Evaluation}\label{pe}
In this section, we evaluate the performance of DLEA. The framework of the simulation model is built by CSIM19 \cite{schwetman2001csim19}, and DLEA is implemented by the C programming together with the MATLAB. Table \ref{table:settings} shows the default parameter settings for  performance evaluation. Unless stated otherwise, the simulations are conducted under these default settings. Specifically, we consider in a $(L \times W)$ $m^2$ road space, where the true locations of vehicles spatially form a Poisson process with the mean arrival rate, $\lambda$, and the mean velocity, $q$. The GPS fix error ($\delta $), as well as the error of the computed inter-vehicle distances ($\varepsilon $), follow the Gaussian distribution.  The default  communication range ($R$) among vehicles is assumed to be 150 meters, which is within the reliable data transmission range supported by the DSRC \cite{ma2012design}.

%In order to quantitatively measure the performance of DLEA, we design the metric of \emph{average location deviation} (ALD), which is calculated by:
%\begin{align*}
%ALD = \frac{{\sum\limits_{v_i  \in V} {\left\| {{\bf{\hat v}}_i  - {\bf{v}}_i } \right\|} }}{{|V|}}
%\end{align*}
%where $\left\| {{\bf{\hat v}}_i  - {\bf{v}}_i } \right\|$ represents the error of the estimated location for $v_i$. Therefore, a small value of ALD %(close to 0) indicates satisfactory performance of DLEA.

\begin {table}
\renewcommand{\arraystretch}{1.3}
\centering \caption{Default settings} \label{table:settings}
\begin{tabular}{|l|l|l|l}
\cline{1-3}
\multicolumn{1}{|c|}{\textbf{Parameter}} & \multicolumn{1}{c|}{\textbf{Default}} & \multicolumn{1}{c|}{\textbf{Descriptions}} &  \\
\cline{1-3}
\multicolumn{1}{|c|}{$L$} & \multicolumn{1}{c|}{500 (m)} & \multicolumn{1}{c|}{length of the road space} &  \\
\cline{1-3}
\multicolumn{1}{|c|}{$W$} & \multicolumn{1}{c|}{9 (m)} & \multicolumn{1}{c|}{width of the road space} &  \\
\multicolumn{1}{|c|}{} & \multicolumn{1}{c|}{} & \multicolumn{1}{c|}{(3 lanes x 3m)} &  \\
\cline{1-3}
\multicolumn{1}{|c|}{$R$} & \multicolumn{1}{c|}{150 (m)} & \multicolumn{1}{c|}{communication range} &  \\
\cline{1-3}
\multicolumn{1}{|c|}{$\lambda$} & \multicolumn{1}{c|}{50 (vehicles/min)} & \multicolumn{1}{c|}{parameter of the Poisson process} &  \\
\cline{1-3}
\multicolumn{1}{|c|}{$q$} & \multicolumn{1}{c|}{50 (km/h)} & \multicolumn{1}{c|}{mean velocity of vehicles} &  \\
\cline{1-3}
\multicolumn{1}{|c|}{$\delta$} & \multicolumn{1}{c|}{$\delta  \sim N(0,10^2 )$} & \multicolumn{1}{c|}{error of GPS fixes} &  \\
\multicolumn{1}{|c|}{} & \multicolumn{1}{c|}{} & \multicolumn{1}{c|}{(Gaussian distribution)} &  \\
\cline{1-3}
\multicolumn{1}{|c|}{$\varepsilon$} & \multicolumn{1}{c|}{$\varepsilon  \sim N(0,1^2 )$} & \multicolumn{1}{c|}{error of computed distances} &  \\
\multicolumn{1}{|c|}{} & \multicolumn{1}{c|}{} & \multicolumn{1}{c|}{(Gaussian distribution)} &  \\
\cline{1-3}
\end{tabular}
\end{table}

\begin{figure}[h]
\subfigure[GPS deviation = 5m]{
\label{dev_5}
\centering
\includegraphics[width=3.1in, height=2.1in]{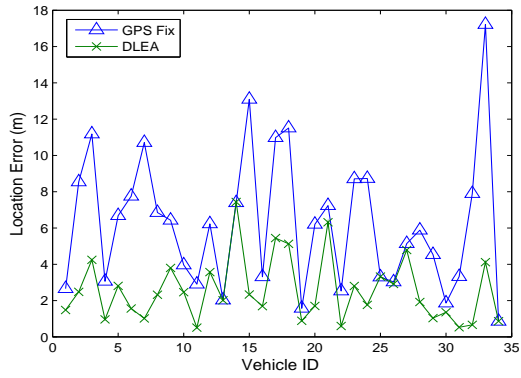}}
\subfigure[GPS deviation = 10m]{
\label{dev_10}
\centering
\includegraphics[width=3.1in, height=2.1in]{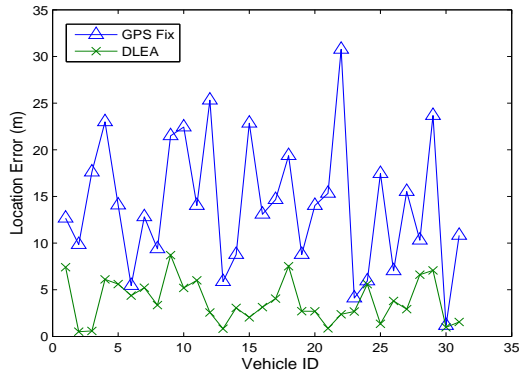}}
\subfigure[GPS deviation = 15m]{
\label{dev_15}
\centering
\includegraphics[width=3.1in, height=2.1in]{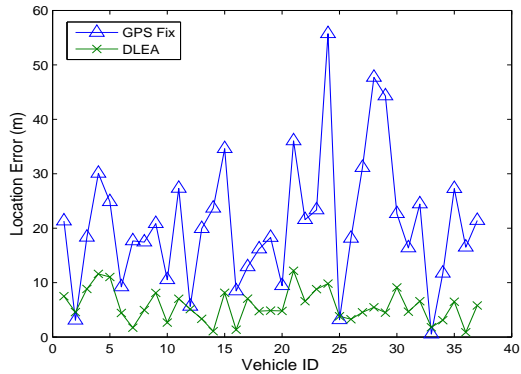}}
\caption{Performance evaluation under different GPS deviation environments}
\label{fig:3}
\end{figure}

As shown in Figure \ref{fig:3}, we evaluate the performance of DLEA under different GPS deviation environments. The x-axis represents the IDs of vehicles along the road. The y-axis represents the location error of each vehicle. In particular, the errors of GPS fixes and DLEA are depicted for comparison, which are calculated by $\left\| {{\bf{\tilde v}}_i  - {\bf{v}}_i } \right\|$  and $\left\| {{\bf{\hat v}}_i  - {\bf{v}}_i } \right\|$, respectively. Clearly, compared with the GPS fixes, the estimated locations by DLEA are much closer to the true locations of vehicles. To give a comprehensive comparison,  Table \ref{table:compare} summarizes the average errors of GPS fixes and DLEA, which are calculated by ${\sum\limits_{v_i  \in V} {\left\| {{\mathbf{\tilde v}}_i  - {\mathbf{v}}_i } \right\|} }/{{|V|}}$ and ${\sum\limits_{v_i  \in V} {\left\| {{\mathbf{\hat v}}_i  - {\mathbf{v}}_i } \right\|} }/{{|V|}}$, respectively. Observed from these statistics, with an increasing value of GPS deviation, the errors of both GPS fixes and DLEA are getting higher. This is reasonable as DLEA estimates the locations based on the original GPS fixes.  Nevertheless,  DLEA always manages to achieve much higher accuracy than GPS fixes across a wide range of conditions.
\begin{table}[h]
\renewcommand{\arraystretch}{1.2}
\renewcommand{\tabcolsep}{0.7cm}
\caption{Average location errors (m)}
\label{table:compare}
\centering
\begin{tabular}{|c|c|c|}
\hline
 & GPS Fix & DLEA \\
\hline
Deviation=5m  & 6.271441 & 2.552837  \\
\hline
Deviation=10m  & 14.119025 &  3.791405 \\
\hline
Deviation=15m  & 20.852989 &  5.678821 \\
\hline
\end{tabular}
\end{table}

\section{Conclusion and Future Work}\label{con}
In this work, motivated by the urgent demand of accurate vehicular localization in safety-critial applications, we proposed a distributed location estimate algorithm, DLEA, to improve the positioning accuracy via cooperative inter-vehicle distance measurement. The implementation of DLEA only relies on  raw GPS pseudorange measurements without the assumption of any particular hardwares or infrastructures, which makes it a potentially inexpensive and practical solution. Besides, DLEA is operated in a distributed fashion, so that it can be adaptable to the highly dynamic vehicular network. Last, we have built the simulation model, and implemented the algorithm for performance evaluation. The simulation results demonstrated that the algorithm can significantly improve the positioning accuracy.

The current solution is effective in scenarios with good GPS signal. In our future work, we will  explore the influence of GPS signal to the positioning performance under different environments, such that we can further refine the solution and make it more robust in compensating and overcoming GPS limitations.

\section*{Acknowledgment}
This research is supported by the Singapore National Research Foundation (NRF) through the Singapore-MIT Alliance for Research and Technology (SMART) Future Urban Mobility (FM) Interdisciplinary Research Group (IRG). We thank Prof. Emilio Frazzoli and Prof. Daniela Rus of MIT for their support and suggestions regarding this work.
\bibliographystyle{IEEEtran}
\bibliography{ref}

% that's all folks
\end{document}